# Creating Power Distribution Network Layouts Using Generative Adversarial Networks and Image-Based Representations

Juan Manuel García-Pérez[1], Carlos Mateo[2]
[1] School of Engineering (ICAI), Comillas Pontifical University, Spain
[2] Institute for Research in Technology (IIT), School of Engineering (ICAI), Comillas Pontifical University, Spain

Corresponding author: Carlos Mateo (cmateo@comillas.edu)

**ABSTRACT:** Utilities increasingly rely on planning and operational tools to cope with the increased penetrations of distributed energy resources, yet the lack of realistic, openly available datasets remains a major barrier for benchmarking and comparison. Traditional test feeders, and recently proposed large-scale synthetic networks alleviate this issue but are typically based on heuristic rules and do not learn directly from data. This paper proposes a generative framework based on Generative Adversarial Networks (GANs) to create power distribution network layouts using image-based representations. The model is trained on rasterised views of distribution systems and can operate in two modes: an unconditional configuration that learns layout patterns from the training dataset, and conditional configurations that incorporate geographical context such as street maps and the spatial distribution of consumers. The methodology includes dataset preparation from Geographic Information System (GIS) sources, GAN architecture design, and the analysis of training stability and image resolution. Results from three representative cases show that the proposed approach can reproduce the topologies of low (LV), medium (MV) and high voltage (HV) feeders and align generated layouts with underlying geographical structures. At the same time, the study reveals limitations related to training stability, resolution-dependent artefacts and limits, and the absence of explicit electrical constraints. The proposed framework constitutes a data-driven complement to existing synthetic network generation methods, and could be applied to propose distribution network layouts for the electrification of new areas. This would require future extensions towards power flow, electrically validated models.

## I. INTRODUCTION

Electric distribution networks are undergoing a profound transformation driven by the rapid deployment of distributed energy resources (DERs), the electrification of other sectors (e.g., electric vehicles and heat pumps), and increasing digitalization. The growing penetration of distributed generation introduces additional uncertainty into distribution networks, which can be mitigated through flexibility provided by distributed storage or flexible loads. This evolution requires enhanced digitalization and the development of new planning and operation tools capable of operating under increasingly uncertainty.

The validation of such tools requires datasets to test and benchmark solutions proposed by different authors. However, real distribution network data is rarely accessible due to confidentiality concerns and the critical infrastructure nature of electrical grids. For decades, this has forced researchers to rely on publicly available benchmark systems such as the IEEE test feeders [1], [2], [3], the PNNL Taxonomy Feeders [4], and the EPRI representative feeders [5]. Although these models have been essential for algorithm comparison and reproducibility, they exhibit several limitations, including reduced scale, limited topological diversity, lack of geospatial information, and simplified operating conditions [6].

In response to these limitations, recent efforts have produced a new generation of large-scale synthetic networks. Among them, the synthetic distribution systems created with the U.S. Reference Network Model (RNM-US) represent a major advancement, synthesizing very large-scale distribution systems and covering several voltage levels [7]. These models have been used to construct a synthetic model of the San Francisco Bay Area [7], as well as a combined transmission-and-distribution model of the state of Texas comprising over 46 million electrical nodes [8]. These developments demonstrate the feasibility of constructing highly detailed, system-wide synthetic datasets for research and validation purposes. Despite their realism, such approaches rely on expert-designed heuristics and rule-based planning algorithms, which limits their ability to directly learn structural patterns from data.

Recent approaches for synthetic distribution system generation include a variety of methods [9], [10], [11], [12]. Shahraeini applies graph models to generate schematics of radial distribution grids [9]. Francisco et al. uses a combination of data collection, manual processes and algorithms (e.g. minimum spanning tree) to generate an Alaskian microgrid synthetic power system model [10]. Lyman et al. focus on the structural validation of synthetic distribution grids using the multiscale flat norm, that is a metric that measures the distance between objects using geometric measure theory [11]. Baecker et al. propose a comprehensive methodological framework, implemented through the pylovo tool, that creates realistic LV synthetic grids [12], including electrical parameters.

In parallel, recent advances in deep learning—particularly in Generative Adversarial Networks (GANs)—have demonstrated remarkable capabilities in synthesizing complex images and structural patterns [13]. GANs comprise a generator–discriminator pair trained using an adversarial approach, enabling the model to learn from data. Their effectiveness has been shown in applications such as image synthesis [14], map generation [15], and structural pattern replication [16]. The work of [17], illustrates the potential of Graph-GAN-based models to learn topological properties of electrical networks through biased random walks, providing synthetic networks whose characteristics closely resemble those of real systems. This demonstrates that generative models can reproduce electrical connectivity patterns and produce networks usable for power-flow studies.

Despite these promising developments, the use of image-based generative models to synthesize distribution network layouts directly from visual and geospatial representations remains largely unexplored. In particular, no deep-learning framework has been proposed that learns spatial distribution system layouts directly from rasterized representations of distribution networks and their geographical context.

This article proposes a generative framework based on GANs for the creation of distribution network topologies. Instead of relying on handcrafted heuristics, the model learns topological layout from existing networks and learns to replicate and reproduce them, offering a data-driven alternative for generating new distribution network layouts. This approach complements existing synthetic models by automatically learning the layout from the data, leveraging on artificial intelligence approaches. A GAN-based generative framework is introduced for proposing distribution network topologies, capable of learning layouts directly from image-based representations. A complete methodological pipeline is presented, covering dataset preparation, model design, training, and evaluation. The approach is validated across several case studies, including multiple voltage levels and different types of geospatial information.

The remainder of this article is organized as follows. Section 2 describes the methodological framework. Section 3 presents the architecture of the generative model and the data-processing pipeline. Section 4 reports the results, and section 5 discusses the limitations of the proposed approach. Finally, section 6 provides the main conclusions, and outlines directions for future research.

## 2. METHODOLOGY

This section presents the methodological framework developed to generate synthetic distribution network layouts using GANs. The approach is fully data-driven and aims to learn directly from image-based representations of electrical distribution systems. The methodology comprises five main components: (i) the overall GAN-based approach, (ii) the model architecture, (iii) dataset preparation, (iv) system variants, and (v) the training strategy and computational setup.

### *A. Overview of the GAN-based Approach*

The proposed method leverages GANs to infer distribution network layouts from image data. Instead of relying on heuristic planning rules, engineering assumptions, or topological constraints explicitly encoded in the model, the GAN learns patterns directly from data examples. Images of real or synthetic networks—representing power lines, consumers, transformer locations, and geographic context—can be used as the main training modality.

Once trained, the model is capable of producing new topologies that reproduce the structural characteristics present in the training data. This enables:

- the generation of synthetic distribution feeders,
- the creation of multiple diverse variants of a given area, and
- the exploration of planning alternatives derived directly from data.

### *B. GAN Architecture: Generator, Discriminator, and Learning Objective*

The model follows a classical GAN architecture, shown in Fig. 1. It consists of two neural networks trained in an adversarial framework:

- **Generator**: Produces synthetic images representing distribution network layouts. Its goal is to generate outputs that resemble realistic feeder layouts similar to those observed in the training dataset.
- **Discriminator**: Receives both input and generated images and learns to distinguish between them. Its feedback drives the generator toward producing more realistic outputs.

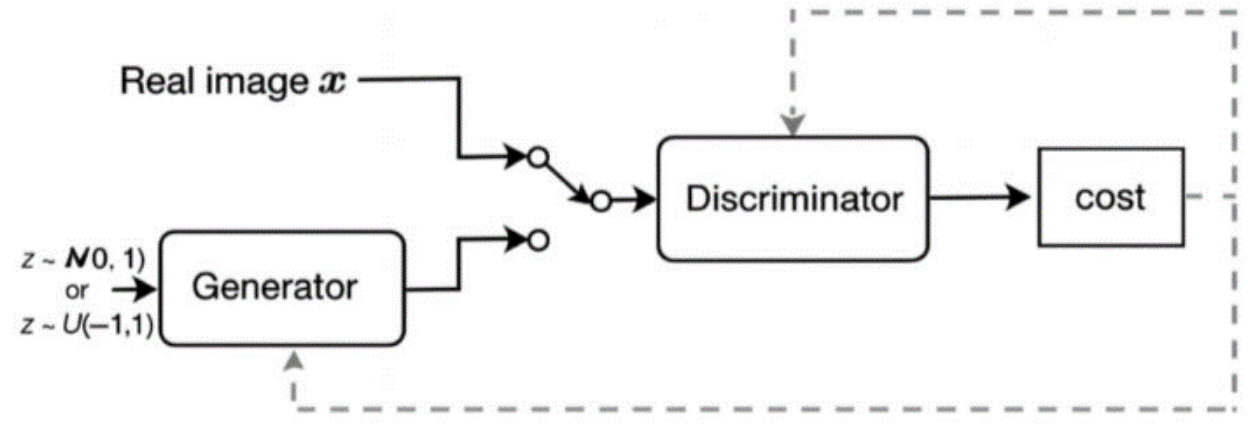

Figure 1. Case 1: Example training images.

Training is formulated as a mini-max optimization problem, where the generator attempts to minimize the discriminator's ability to detect synthetic samples, while the discriminator attempts to maximize it. Standard adversarial loss functions are used [13], combined with stabilization techniques (e.g., gradient penalty, batch normalization, image downsampling) to improve convergence and limit mode collapse.

Depending on the experimental setup, the generator may operate:

- unconditionally (without contextual input images), producing generic distribution network layouts; or
- conditionally (with auxiliary geographical context). In this set-up, input images such as the street map and the consumers of the area being planned, are provided.

The architectural backbone is convolutional, enabling the model to capture patterns in the images.

### C. Dataset Preparation

The training dataset consists of image-based representations of distribution systems extracted from existing synthetic distribution system datasets [7]. The data preparation pipeline includes:

#### 1) EXTRACTION OF TRAINING SAMPLES

Network layouts are exported as raster images from GIS-based models. These images encode the layout of power lines, the location of buses and substations, and other relevant elements. The dataset covers several voltage levels and street-map configurations.

#### 2) PREPROCESSING AND IMAGE SEGMENTATION

Each image undergoes a preprocessing pipeline that includes:

- generating images from GIS data with a normalized resolution,
- removing irrelevant areas (e.g. areas with few consumers),
- line width adjusting or coloring when needed (e.g. to represent phases, voltage levels, and transformers), and
- resizing to a fixed dimension.

The objective is to propose a raster image representation of distribution systems, ensure consistency across training samples, and reduce noise introduced by empty regions.

#### 3) DATASET STRUCTURING

Images are grouped by scenarios: unconditional (without geographical context) or conditional (with geographical context), and voltage levels (low, medium or high-voltage layouts).

### D. System Variants

Two variants of the generative system are considered:

- **Unconditional GAN**: It produces synthetic distribution network layouts without explicit contextual inputs. The model learns general spatial layout patterns from the training dataset and generates plausible generic feeder layouts. This configuration is suitable for creating diverse synthetic distribution network layouts for benchmarking and analysis.
- **Conditional GAN** (using auxiliary input context): With this configuration, the generator receives additional contextual images (e.g., street maps and consumer distributions) and produces a compatible network layout aligned with the provided geographical information. This variant enables the generation of layouts tailored to specific areas and supports early-stage exploration of geographically consistent network layouts to be used for the electrification of new areas.

Both variants use the same core GAN architecture but differ in the structure of the generator's input layer.

### E. Training Strategy and Computational Setup

Training is performed using mini-batch stochastic gradient descent. Key components of the training strategy include:

- Adversarial training: Generator and discriminator are updated alternately to preserve training balance.
- Learning-rate tuning: Separate learning rates are used for both networks to stabilize the adversarial dynamics.
- Epoch scheduling: Experiments indicate that stability improves as the number of epochs increases, although training may degrade if extended excessively.
- Stabilization mechanisms: Techniques commonly used in GAN training, such as batch normalization and careful control of training duration, are employed to improve convergence and reduce instability effects such as mode collapse.
- Hardware: Training is carried out using GPU acceleration to handle the computational load associated with high-resolution images and deep convolutional layers.

Trained models are saved periodically, and intermediate outputs are inspected to assess convergence and training stability.

## 3. Architecture of the Model and Generation Pipeline

This section describes the structure of the generative model and the workflow used to produce synthetic distribution network layouts. The model follows a classical Generative Adversarial Network (GAN) design, adapted to operate on image-based representations of distribution networks.

### A. Structure of the Generator

The generator is implemented as a convolutional neural network whose role is to transform either a latent noise vector (in the unconditional configuration) or an input image containing geographical context (in the conditional configuration) into a synthetic distribution network layout. The generator consists of:

- an input stage, receiving either random noise or contextual images such as street layouts, the spatial distribution of consumers, or lower-voltage network layouts;
- a sequence of convolutional and upsampling layers that progressively construct spatial features;
- nonlinear activation functions (ReLU); and
- a final convolutional layer producing the generated image encoding the network topology.

In the conditional configuration, contextual images are fed directly into the generator. The model learns how the geographical context influences the resulting electrical layout purely from example data.

### B. Structure of the Discriminator

The discriminator is a convolutional neural network trained to distinguish between real images and those produced by the generator. Its architecture includes:

- convolutional blocks that reduce spatial resolution while extracting higher-level features;
- nonlinear activations functions (LeakyReLU);
- strided convolutions for downsampling; and
- a final output layer providing a scalar value indicating whether the input is real or generated.

The discriminator's feedback drives the optimization of both networks within the adversarial training framework.

### C. Use of Street Maps and Consumers as Input

In the conditional configuration, the generator receives additional image layers providing geographical and network-related context for the area under study. These inputs include:

- the spatial distribution of consumers
- street map, and
- low voltage network layouts, used as contextual information for generating medium- or high-voltage network layouts.

These contextual images are introduced directly as an input, allowing the generator to produce network layouts that reflect spatial structures present in the source data. The model learns statistical associations between contextual patterns and the resulting network layouts without explicit electrical or planning constraints.

### D. Training Behaviour and Convergence Indicators

Training progress is assessed mainly through qualitative inspection and the stability of adversarial dynamics. Key indicators used to assess convergence include:

- the progressive emergence of coherent layouts,
- reduction of background noise and visual artefacts,
- diversity of generated samples across training iterations,
- continuity in line segments and recognizable spatial patterns, and
- alignment between generated layouts and contextual inputs in conditional scenarios.

Training instabilities may appear when training is prolonged for an excessive number of epochs or when the discriminator becomes overly dominant, leading to degradation in the visual quality and structural coherence of generated layouts. These observations inform the selection of training duration and update scheduling.

### E. Generation Pipeline

The generation process follows a consistent workflow:

1) Input Preparation
    a. Latent noise vector (unconditional generation)
    b. Contextual image layers (conditional generation)
2) Generator Forward Pass
   Produces a synthetic network layout at the specified image resolution.
3) Discriminator Evaluation
   Compares real and generated images to provide adversarial feedback.
4) Weight Updates
   Both networks are updated alternately according to the adversarial loss.
5) Output Selection and Inspection
   Periodic samples are stored to monitor the evolution of visual quality and structural coherence during training.

This pipeline is iterated until the generated layouts exhibit stable visual quality and spatial consistency.

## 4. Results

This section presents the results obtained from three experimental scenarios, each exploring a different configuration of the proposed generative model. The three cases illustrate an unconditional configuration (Case 1) and two conditional configurations (Cases 2 and 3) applied to different voltage levels.

- **Case 1** explores unconditional generation and reproduces generic feeder layouts using only statistical information learned from the training dataset.
- **Case 2** focuses on low-voltage network layouts conditioned on street maps and consumer locations.

- **Case 3** generates medium/high-voltage layouts considering the low voltage network. This aims to replicate how the U.S. Reference Network Model plans distribution systems using a bottom-up approach, starting with the lower voltage levels, and proceeding upstream [7].

All cases rely on raster image-based representations of distribution networks during training, differing only in the type and presence of contextual input information.

### *A. Case 1 — Unconditional GAN*

#### 1) INPUT CONFIGURATION

Case 1 corresponds to the scenario in which the generator receives no contextual images. The model operates exclusively from a latent noise vector and learns to reproduce spatial layout patterns present in the training dataset. Example training images are shown in Fig. 2, where red, green and blue lines represent power lines with phases A, B & C, and consumers are represented as small dots.

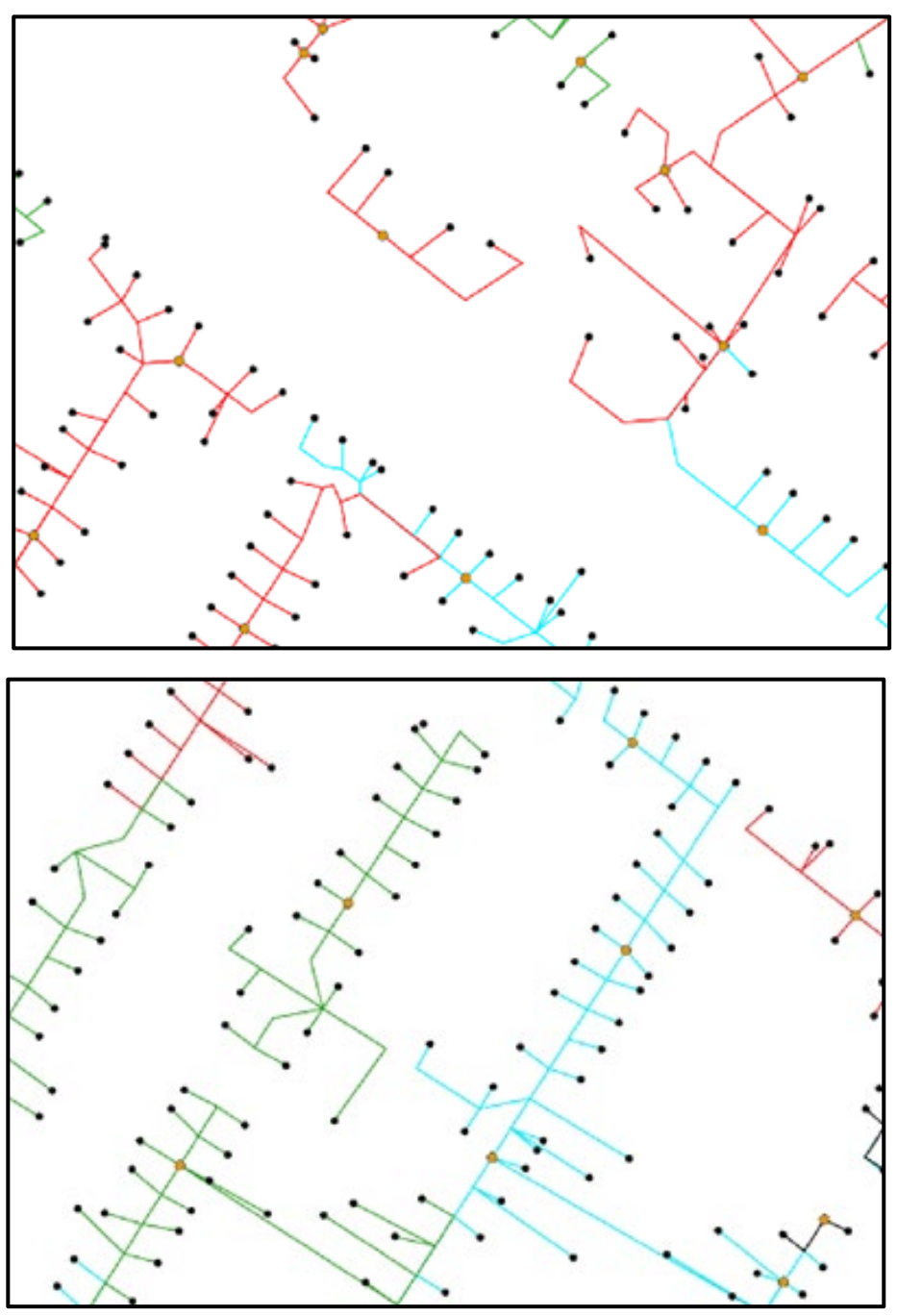

Figure 2. Case 1: Example training images.

#### 2) TRAINING BEHAVIOUR

The evolution of the generated output over the training is illustrated in Fig. 3. Training begins with highly noisy, fragmented outputs. With progression through epochs:

- line segments become longer and more coherent,
- branching structures start to emerge, and
- the overall layout increasingly resembles a distribution feeder.

Training stability is strongly influenced by the balance between the generator and the discriminator. When the discriminator becomes overly dominant, sudden degradations in visual quality are observed. These effects are mitigated by adjusting the relative update frequency of both networks.

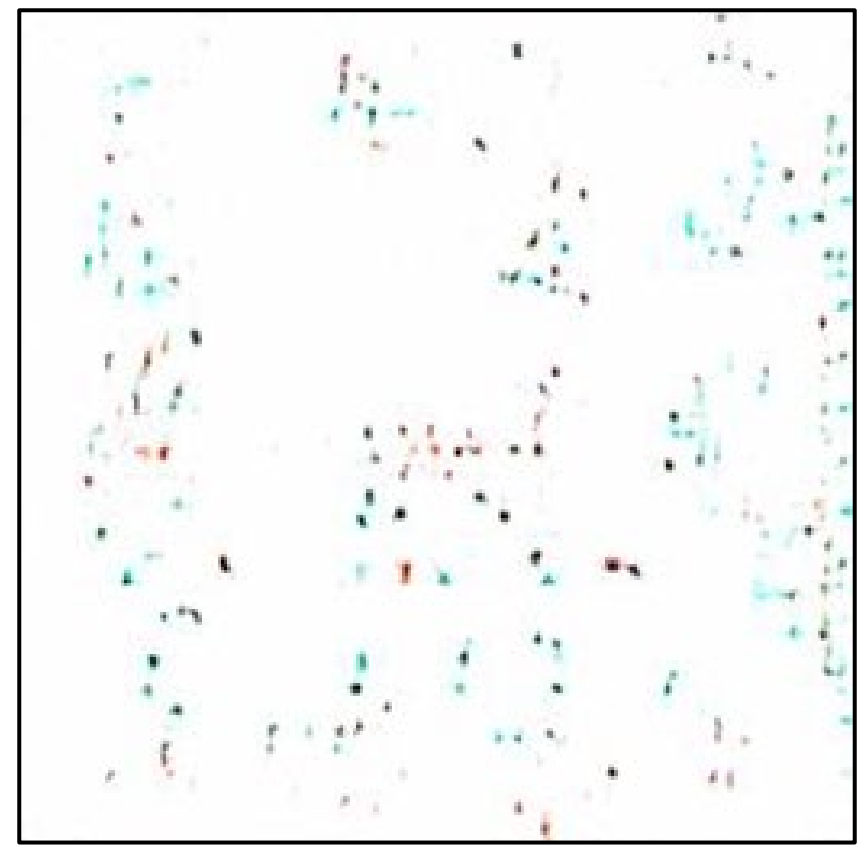

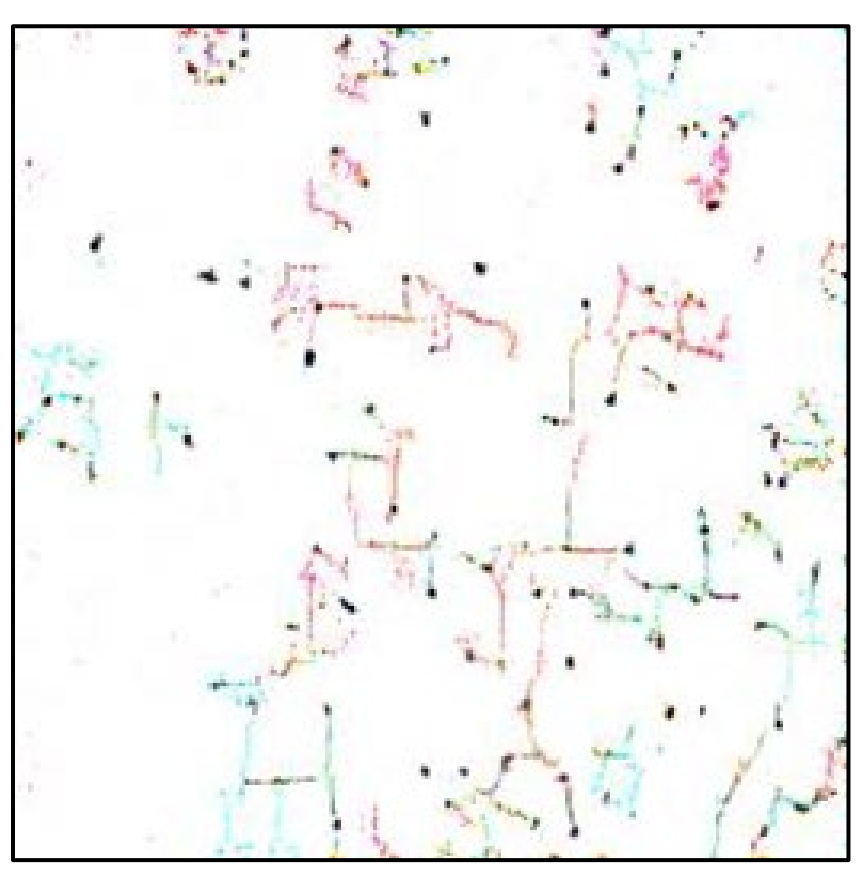

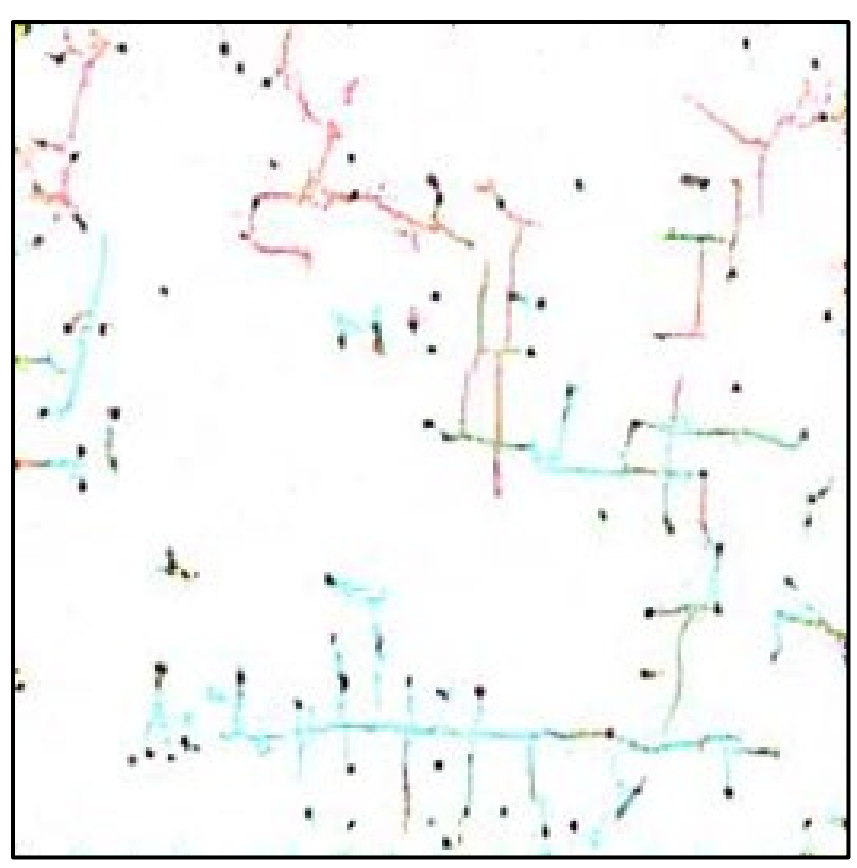

Figure 3. Case 1: Evolution of the generated output over the training (top: initial generated inputs, bottom: final generated outputs)

#### 3) GENERATED TOPOLOGY

The final outputs (right of Figure 2), exhibit clear radial and branching patterns commonly observed in distribution network layouts. Even in the absence of contextual information, the model learns to reproduce global spatial characteristics such as main feeder paths and secondary branching. However, distortions and irregularities remain, reflecting the limitations of unconditional generation. This case serves as a baseline for comparison with the conditional scenarios.

### *B. Case 2 — Low-voltage topology conditioned on street map and consumer distribution*

**1)** INPUT CONFIGURATION

Case 2 incorporates contextual information:

- the street layout of the area, and
- the spatial distribution of consumers.

This configuration reflects a typical low-voltage supply environment, where feeders must connect a large number of closely spaced demand points, while taking into account the existing street map.

2) TRAINING BEHAVIOUR

Compared to Case 1, a clear improvement in training stability and output coherence is observed. Generated images exhibit more consistent line continuity, reduced background noise, and clearer structural organization. This case highlights the importance of geographical context in guiding the generation process and improving the spatial plausibility of the resulting layouts.

3) GENERATED TOPOLOGY

The final generated outputs, presented in Fig. 4, shows:

- dense branching patterns characteristic of low-voltage distribution networks,
- alignment between feeders and street orientation, and
- clear correspondence between consumer clusters and local network structure.

These results demonstrate that the conditional GAN successfully integrates multiple contextual layers to produce spatially consistent low-voltage network layouts.

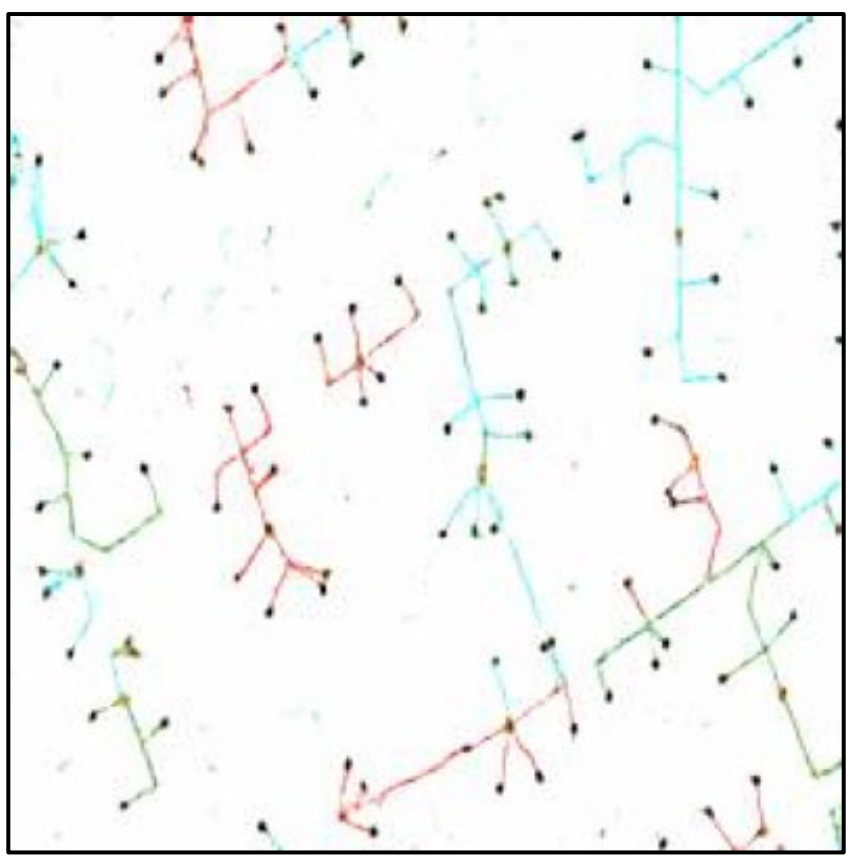

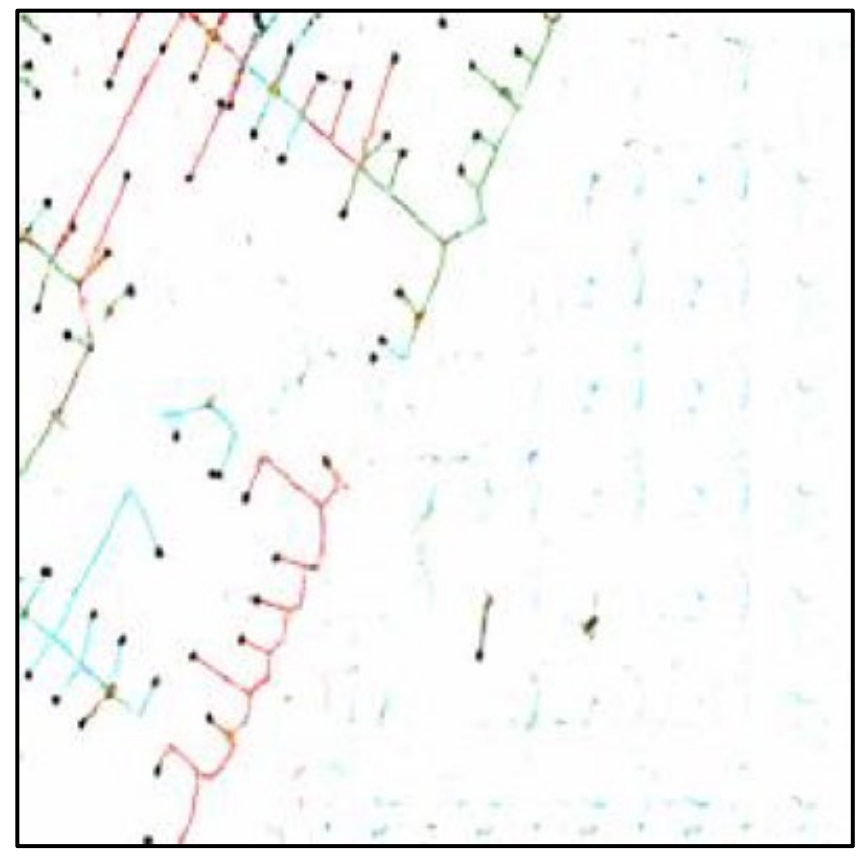

Figure 4. Case 2: Examples of generated outputs

### *C. 4.3. Case 3 — Medium- or high-voltage topology conditioned on low voltage*

1) INPUT CONFIGURATION

In Case 3, the low-voltage network layout is provided as contextual input to guide the generation of medium- and higher-voltage distribution layouts. The objective is to explore whether the model can learn spatial relationships between different voltage levels in a data-driven manner.

2) TRAINING BEHAVIOUR

Training experiments indicate that extending training beyond approximately 350 epochs does not lead to further improvements in output quality. In some cases, prolonged training results in increased background noise and visual degradation, suggesting overfitting. These observations highlight the importance of monitoring training duration and interrupting training when degradation effects appear.

3) GENERATED TOPOLOGY

The generated layouts are shown in Fig. 4. Red, green and blue lines represent power lines with phases A, B & C, while black lines represent three-phase power lines. The results exhibit coherent upstream structures that align spatially with the underlying low-voltage network. Line thickness variations reflect different voltage levels in the image representation, and the generated layouts preserve the general spatial organization imposed by the low-voltage input. These results suggest that the model is capable of learning contextual spatial relationships between voltage levels, despite the absence of explicit electrical constraints.

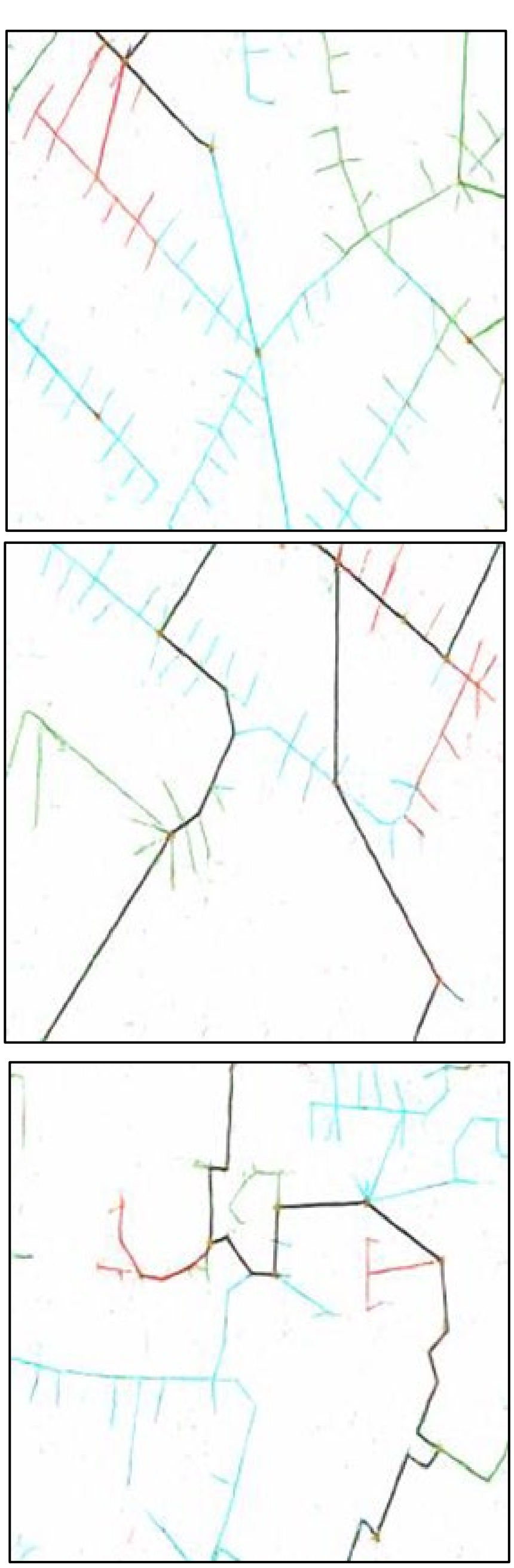

Figure 4. Case 3: Examples of generated outputs

## 5. DISCUSSION AND LIMITATIONS

The experiments conducted across the three cases highlight both the potential and the current limitations of using GAN-based models to generate image-based distribution network topologies. The discussion below summarizes the main observations regarding model behaviour, training stability, scalability, and dependence on the input data.

### A. Training Stability and Convergence Challenges

GAN training proved highly sensitive to the balance between the generator and the discriminator. Several forms of training instability were observed during the experiments, including:

- Gradient explosion, often occurring when the discriminator becomes too dominant.
- Sudden degradation of output quality, which appears intermittently during training.
- Mode collapse, in which the generator produces highly similar images across iterations.
- Noise artefacts, especially at early epochs or after prolonged training.

Achieving stable convergence required careful selection of the number of training epochs and adjustment of the relative update frequency between the generator and the discriminator. Excessive training often led to degradation in visual quality, highlighting the need for improved mechanisms to regulate adversarial dynamics.

### B. Sensitivity to Image Resolution

The quality of the generated layouts depends strongly on the chosen resolution. Higher resolutions allow the model to design larger distribution areas details, but they also:

- increase training instability,
- amplify the risk of generating noisy or incomplete segments, and
- require greater computational resources.

Lower resolutions produce smoother and more stable results but limit the spatial scale that can be represented. This trade-off is particularly relevant for medium- and high-voltage distribution layouts (Case 3), where larger spatial extents would be required.

### C. Dependence on the Input Dataset

The generative performance is constrained by the characteristics of the training dataset:

- the model reproduces only the spatial patterns present in the available images,
- areas with regular street layouts are easier for the model to learn,
- sparse or irregular regions lead to more variability and artefacts in the output, and
- the generator may overfit recurring structures if the dataset is not sufficiently diverse.

This dependence is especially evident in unconditional generation, where the absence of contextual input limits the model to the statistical average of the dataset. Conditional cases alleviate this to some extent but still reflect the biases inherent in the input data.

### D. Complexity of Low-Voltage Scenarios

Low-voltage networks present a higher degree of structural variability and branching density than medium- or high-voltage networks. The experiments show that:

- the generator requires more epochs to learn LV patterns,

- small distortions are more noticeable due to the density of segments, and
- integrating multiple contextual layers increases training difficulty.

These factors make low-voltage topology generation a more challenging task, requiring additional training strategies or architectural refinements to improve robustness.

### *E. Limitations of Image-Based Representation*

Using images as the primary representation introduces intrinsic limitations:

- pixel-level artefacts can disrupt the continuity of power lines, and
- the model cannot explicitly enforce electrical constraints such as radiality, connectivity, or feeder capacity.

While the generated images capture spatial and visual patterns of distribution networks, they do not directly encode electrical properties. Any conversion from image-based layouts to electrically valid network models would require additional post-processing steps that are beyond the scope of the present work.

## 6. CONCLUSIONS

This work presents a generative framework based on Generative Adversarial Networks for the creation of synthetic distribution network topologies using raster image-based representations. The proposed approach learns structural patterns directly from data and produces layouts that reflect the spatial characteristics of real or synthetic distribution areas. The model is capable of operating both without contextual information—relying solely on learned statistical features—and with multiple forms of geographic context, such as street layouts and consumer distributions.

The results show that the method can generate low, medium and high voltage feeder topologies that resemble those present in the training dataset. In the paper, GANs have been trained using a publicly available dataset [7], but the methodology does not depend on the dataset and would be applicable to actual networks, which would require access to data from a utility. When contextual images are provided, the generator successfully aligns the topology with the underlying contextual image and, in the case of low-voltage scenarios, adapts branching density to consumer density. These findings confirm that conditional inputs significantly enhance the realism and spatial coherence of the generated layouts compared with unconditional generation. A conditional configuration, allows the model to be used not only for building synthetic networks, but also to propose network layouts for electrifying new areas, taking into account information such as the street map of the area, or the consumers to be supplied.

The study also highlights several challenges. Training stability remains a major limitation, as the adversarial process is sensitive to the balance between generator and discriminator. The model shows vulnerability to noise, mode collapse, and resolution-dependent artefacts, particularly in high-detail scenarios. Furthermore, the image-based representation does not incorporate electrical constraints such as connectivity or radiality, which would be necessary for producing fully operational network models.

Despite these limitations, the proposed approach demonstrates that generative models can effectively capture and replicate spatial patterns in distribution networks. The method provides a data-driven alternative to heuristic-based synthetic network generation and offers a foundation upon which more advanced models can be developed. Its ability to produce diverse and geographically consistent layouts positions GAN-based generation as a promising tool for expanding the set of available test feeders used in distribution network research. The fact that this methodology could be applied to actual networks, would be a clear advantage over heuristics, replacing hard-coded heuristic design solutions, with actual designs directly learnt from data, which is particularly interesting to propose initial topological designs in order to electrify new areas. For this application, it would also be interesting the ability of GANs to propose several candidate solutions for the same area, that could be analyzed and benchmarked.

Looking ahead, future work should focus on extending the image-based approach toward integrating electrical parameters such as impedances, equipment size and load profiles, exploring other energy sectors, and evaluating operational performance through established power flow simulation tools. Scaling the methodology to larger regions and using training datasets of actual networks would further enhance its applicability.

## AI DECLARATION

An artificial intelligence tool was used to assist in translating and summarizing a longer Spanish version of this study. The authors guided and reviewed the process, and edited the final manuscript.